\documentclass[smallextended]{svjour3}       % onecolumn (second format)
\smartqed  % flush right qed marks, e.g. at end of proof
\usepackage{amsmath}
\usepackage{graphicx}
\usepackage[utf8]{inputenc}

\usepackage[hyphens]{url} % not crucial - just used below for the URL
\usepackage{xcolor}
\usepackage{scrextend}
\usepackage[colorlinks=true, 
pdfstartview=FitV, 
bookmarks=true, 
bookmarksnumbered=true, 
breaklinks]{hyperref}
\usepackage{color}
\definecolor{blue}{rgb}{0.0, 0.0, 1.0}
\definecolor{red}{rgb}{1.0, 0.0, 0.0}
\definecolor{royalblue}{rgb}{0.0, 0.14, 0.4}
\hypersetup{linkcolor=royalblue, 
	citecolor=blue, 
	urlcolor=royalblue}

\def\orcid#1{\kern .08em\href{https://orcid.org/#1}{\includegraphics[keepaspectratio,width=0.7em]{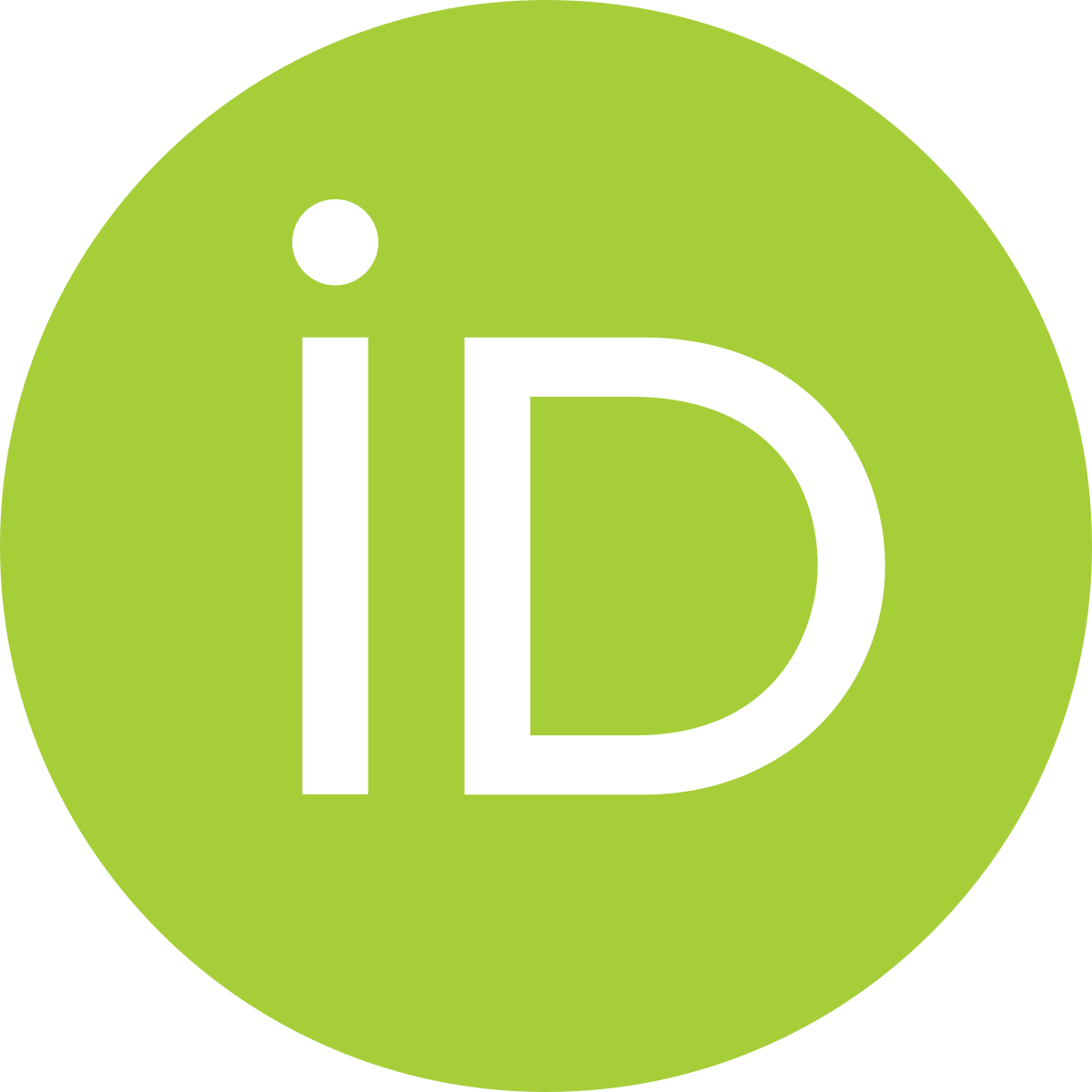}}}

\begin{document}

\title{Classifying near-threshold enhancement using deep neural network}
%\subtitle{Do you have a subtitle? If so, write it here} 

%    \titlerunning{Classifying near-threshold enhancement}

\author{Denny Lane B. Sombillo \and  Yoichi Ikeda \and  Toru Sato \and Atsushi Hosaka}

    \authorrunning{DLB Sombillo, Y Ikeda, T Sato, A Hosaka}

\institute{
        Denny Lane B. Sombillo\orcid{0000-0001-9357-7236} \at
     National Institute of Physics, University of the Philippines Diliman, Quezon City 1101, Philippines \\
     \email{\href{dbsombillo@up.edu.ph}{\nolinkurl{dbsombillo@up.edu.ph}}}  %  \\
    \and
        Yoichi Ikeda\orcid{0000-0002-2235-1464} \at
     Department of Physics, Kyushu University, Fukuoka 819-0395, Japan \\
     %\email{\href{mailto:djf@wef}{\nolinkurl{djf@wef}}}  %  \\
    \and
    Denny Lane B. Sombillo\orcid{0000-0001-9357-7236} 
    \and Toru Sato\orcid{0000-0001-5216-5657} 
    \and Atsushi Hosaka\orcid{0000-0003-3623-6667} 
    \at
    Research Center for Nuclear Physics (RCNP), Osaka University, Ibaraki, Osaka 567-0047, Japan 
    \\
    \email{\href{sombillo@rcnp.osaka-u.ac.jp}{\nolinkurl{sombillo@rcnp.osaka-u.ac.jp}}} 
	}

\date{Received: date / Accepted: date}

\maketitle

\begin{abstract}
	One of the main issues in hadron spectroscopy is to identify the origin of threshold or near-threshold enhancement. Prior to our study, there is no straightforward way of distinguishing even the lowest channel threshold-enhancement of the nucleon-nucleon system using only the cross-sections. The difficulty lies in the proximity of either a bound or virtual state pole to the threshold which creates an almost identical structure in the scattering region. Identifying the nature of the pole causing the enhancement falls under the general classification problem and supervised machine learning using a feed-forward neural network is known to excel in this task. In this study, we discuss the basic idea behind deep neural network and how it can be used to identify the nature of the pole causing the enhancement. The applicability of the trained network can be explored by using an exact separable potential model to generate a validation dataset. We find that within some acceptable range of the cut-off parameter, the neural network gives high accuracy of inference. The result also reveals the important role played by the background singularities in the training dataset. Finally, we apply the method to nucleon-nucleon scattering data and show that the network was able to give the correct nature of pole, i.e. virtual pole for ${}^1S_0$ partial cross-section and bound state pole for ${}^3S_0$.
\\

%    \subclass{
%                    MSC code 1 \and
%                    MSC code 2 \and
%            }

\end{abstract}

\def\spacingset#1{\renewcommand{\baselinestretch}%
{#1}\small\normalsize} \spacingset{1}

\hypertarget{intro}{%
\section{Introduction}\label{intro}}
\hspace{\parindent}
Recently discovered near-threshold phenomena open a renewed interest in the study of hadron spectroscopy~\cite{Olsen2018,Pc2019,dijpsi2020,chi3872}. Some studies suggest that a two-hadron bound state may cause the observed near-threshold enhancements, similar to the case of the deuteron~\cite{Hyodo2013,Yamaguchu2020,Haidenbauer2021}. However, if the attractive potential is not strong enough, a virtual state may also cause the same enhancement~\cite{Guo2018,Threshold2021}. Another possibility is purely kinematical, where a threshold cusp emerges due to two-hadron rescattering~\cite{TriangleAndCusp}. The threshold cusp can be enhanced if there is a nearby virtual state pole in the relevant threshold. With the limited energy resolution and the error bars in the experimental data, both the dynamical and kinematical explanations of near-threshold enhancements are equally possible. It is, therefore, crucial to formulate a method that can distinguish different possibilities.

\begin{figure}
	\centering
	\includegraphics[width=0.9\columnwidth]{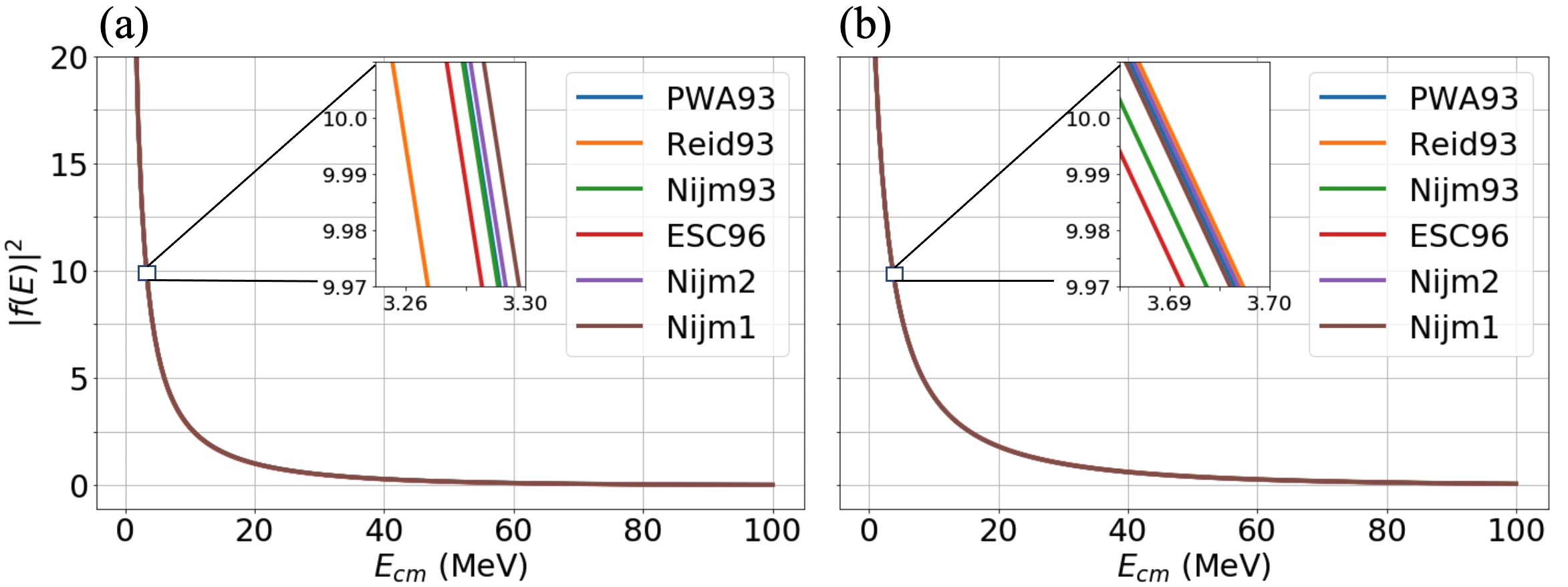}
	\caption{S-wave partial cross-sections of nucleon-nucleon in spin singlet (a) and spin triplet (b) configurations. The data are obtained from the different potential models and partial wave analyses in~\cite{NNonline}. The insets show the small difference among the cross-sections.}
	\label{fig:nn_scatt}
\end{figure}

In this paper, we show how a deep neural network (DNN) can differentiate a bound and a virtual state threshold enhancement in the scattering cross-section. For a more concrete demonstration we use the nucleon-nucleon partial cross-sections as our experimental data. There are six cross-sections that can be obtained using partial-wave analyses and potential models in Ref.~\cite{NNonline}. These are shown in Figure~\ref{fig:nn_scatt} where (a) corresponds to spin-singlet and (b) for the spin-triplet. Note that both the spin-singlet and spin-triplet cross-sections exhibit large values at the threshold. It is a common belief that without knowing the properties of deuteron, it is impossible to distinguish the two enhancements using only the cross-sections. The nucleon-nucleon cross-sections is a perfect example to simulate our ignorance of near-threshold resonances in coupled-channel problems. Unlike the deuteron, hadron resonances are short-lived, and the only way of inferring their existence and properties is through the scattering experiment. In this study, our objective is to distinguish the two enhancement structures without using the known properties of deuteron.

The problem of identifying the nature of enhancement can be reduced to a classification task. That is, given a collection of s-wave scattering cross-sections, we want to tell if the observed enhancement is caused by a physical state. One promising tool to address the classification problem related to threshold phenomena is deep learning~\cite{Alpaydin2010}. There is already a growing trend of using data science approach in the study of physical sciences~\cite{MLandPS}. In the particle physics, machine learning are usually employed in the experimental level where certain signatures of known particles are used to automate particle identification and classify events~\cite{MLandParticle}. In~\cite{Sombillo2020}, we start to extend the applicability of deep learning to analyze the final scattering data. Our first attempt is to distinguish the general structures observed in the single-channel scattering cross-sections which includes resonances or threshold enhancements due to bound or virtual state. Recently, we designed a DNN that can be used to probe the pole structure of coupled-channel scattering amplitude~\cite{Sombillo2021}. The main objective of our project is to simulate the analytic structure of S-matrix in the generation of training dataset and draw information from the experimental data with less bias as possible during the DNN inference stage. In the present study, we focus only on distinguishing bound and virtual enhancements at the threshold and exploring different DNN designs. 

We take the following steps. First, we use the general properties of S-matrix to generate a collection of simulated cross-sections. Here, the collections of cross-sections need not contain the experimental data. Then, we considered six designs of DNN and monitor their learning performance against the training dataset. To ensure that the trained DNN can be applied to dataset outside the training, we prepare a separate set of cross-sections using the separable potential. Finally, we apply the trained and validated models using the nucleon-nucleon data.

\hypertarget{smat}{%
	\section{Generation of teaching dataset}\label{smat}}
\hspace{\parindent}
It is impossible to know the exact form of S-matrix due to the non-perturbative nature of QCD. However, we can simulate the 
analytic structure of S-matrix using the general properties such as analyticity, unitarity, and hermiticity. First, causality, i.e., the scattering cannot precede the collision, implies that the scattering amplitude should be analytic in the first quadrant of the complex momentum plane. Imposing analyticity on the S-matrix means that we cannot have singularities on the upper half of the complex momentum plane, except along the imaginary axis. This restriction creates an unbalanced distribution of singularities on the complex momentum plane which turn out to be useful in distinguishing structures above the threshold. Next is that the probability conservation requires that the S-matrix be unitary. Finally, hermiticity requires that the S-matrix is real-valued below the lowest threshold. These properties and our freedom to put the singularities anywhere, as long as they are consistent with analyticity, will help us construct a set of physically realizable cross-sections. 

We consider two types of background singularities: one that can produce a left-hand branch-cut with adjustable threshold and the other one is a simple background pole. The following parametrization will suffice:
\begin{equation}
	S(p)=\exp\left[2i\eta \tan^{-1}\left(\dfrac{p}{\Lambda}\right)\right]
	\left(\dfrac{p+i\gamma_{far}}{p-i\gamma_{far}}\right)
	\left(\dfrac{p+i\gamma_{near}}{p-i\gamma_{near}}\right)
	\label{eq:smat}
\end{equation}
where the first factor containing $\eta,\Lambda$ parameters is the background simulating left hand cut in the scattering energy plane. At this point, non-relativistic treatment will suffice because we are only concern on the structures near the threshold. However, for a more general coupled-channel case, the background introduced here is useful to simulate the relativistic left-hand cut. Now, the $\eta$ can be set to produce an attractive or repulsive background while the $\Lambda$ adjust the location of left-hand threshold. The second factor simulates a simple pole background through the parameter $\gamma_{far}$. The third factor is responsible to the nature of threshold enhancement in the simulated cross-sections. If $\gamma_{near}>0$, we get a bound state pole enhancement. Otherwise, it is a virtual state enhancement. The relevant scattering cross-section is obtained using the relation $|f(p)|^2=\left|{[S(p)-1]}/{(2ip)}\right|^2$.

Notice that it will be impossible to distinguish a virtual pole enhancement with a bound state in the absence of background singularity. 
Specifically, the cross-sections become identical, i.e., $|f(p)|^2=1/(p^2+\gamma_{near}^2)$, whether the $\gamma_{near}$ is negative or positive.
This difficulty is remedied by simulating the analytic structure of S-matrix where various background singularities are introduced.
We can, therefore, expect that there will be some differences between virtual and bound threshold-enhancements. 

All the cross-sections are produced with known signs of $\gamma_{near}$, allowing us to label them accordingly.
We generate a sizable training dataset by choosing random values of all the parameters in Eq.~\ref{eq:smat}. Overall, we produced $4\times 10^6$ labeled cross-sections. We only use $3.2\times 10^6$ of these labeled cross sections for the direct training and hide the remaining $8\times 10^5$ for the testing. 

\hypertarget{dnn}{%
	\section{Development of deep neural network models}\label{dnn}}
\hspace{\parindent}
The goal of the DNN is to establish a map between the input space of cross-sections and the output bound-virtual classification space. There is no definite rule on how to construct the optimal DNN but there are already good practices that we can imitate. A basic DNN consists of input layer, some number of hidden layers, and an output layer. The nodes of the input layer are numerical values representing the features of the input data. For our purpose, these input node values are cross-sections on the equal-spaced energy region ($0-100$ MeV). For a classification task, it is optimal to use the ReLU (rectified linear unit) as the activation function of the hidden-layer nodes and softmax activation for the output nodes~\cite{Aggarwal2018}. 

The basic operation of a DNN is to take the input numerical values and form a linear combination using the weights and biases. The result of linearly transformed input are fed to the node of the next layer. The activation function will then assign a numerical value to that node and a new linear combination is formed to be passed to the next nodes of the next layer. The combinations of linear transformation and the non-linear effect of activation functions allow the DNN to map the input space to the output space. The mapping can only be successful if the weights and biases are optimized. This can be done in the training stage.

During the training, the present value of cost-function is estimated using the forward pass. Here, the simulated cross-sections are fed to the DNN and the output are compared to the actual labels of the inputs. The discrepancy between the input and output labels are used to estimate the cost-function.
Usually, we use the mini-batch system to add stochasticity to the estimated cost. Now, the values of the weights and biases are updated using gradient descent. This updates happen in the backward manner where the partial derivatives of the cost-functions are calculated. The updating part is referred to as the backpropagation. One complete forward pass and backpropagation of all the simulated cross-sections corresponds to one training epoch. 

\begin{figure}
	\centering
	\includegraphics[width=\columnwidth]{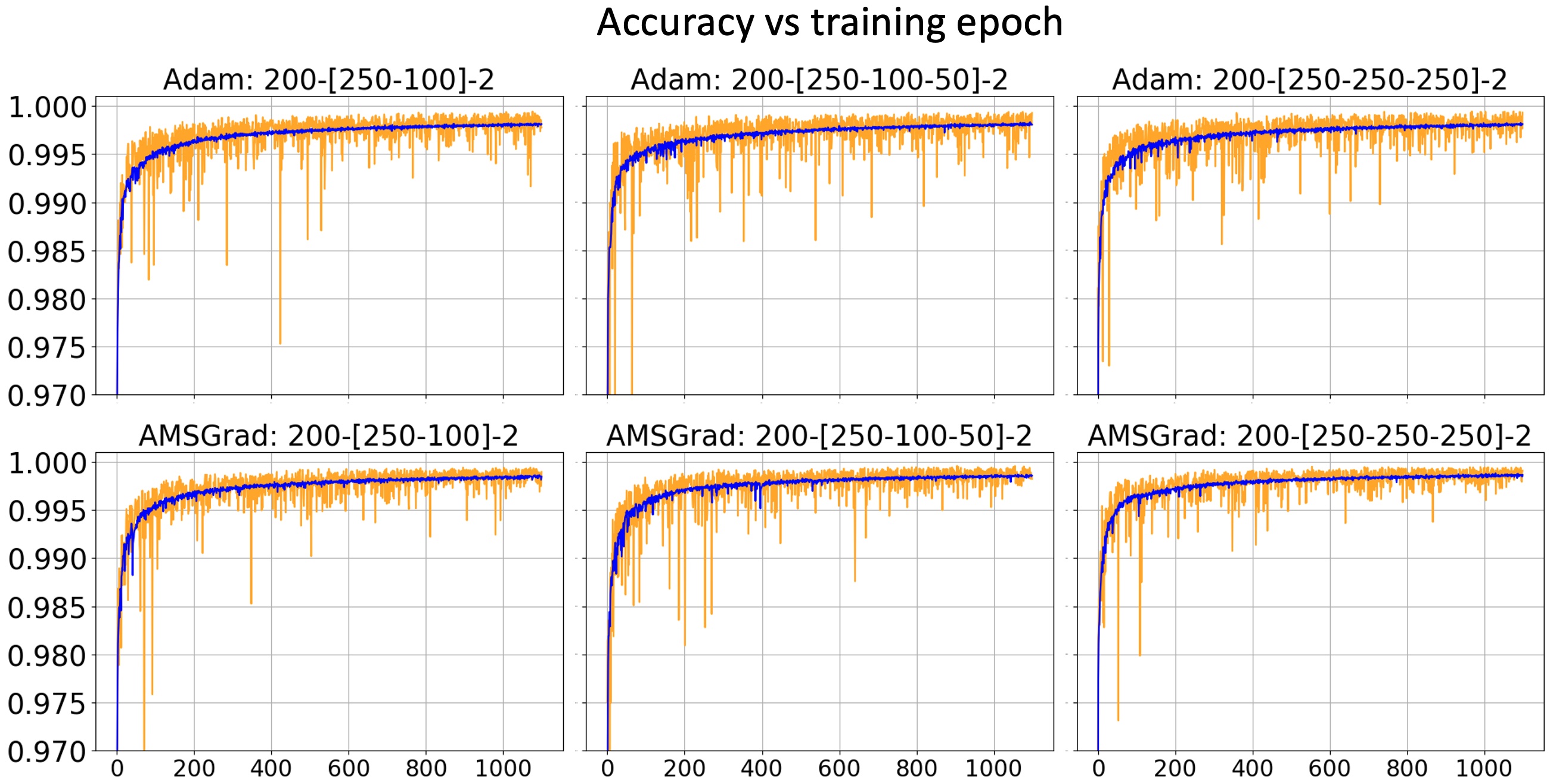}
	\caption{Training performance of all the DNN models. The blue lines are for the training dataset and the orange lines are for the testing set.}
	\label{fig:training}
\end{figure}

In this study we designed six DNN models using three different hidden-layer architectures and two types of optimizers.
The architectures are labeled according to the notation: 200-[XXX-...-XXX]-2 where $200$ and $2$ are the numbers of input and output nodes, respectively. The numbers inside the square bracket give the number of nodes in each hidden layer separated by ``-" for each layer. The optimizers used to perform the backpropagation are Adam and AMSGrad. The Adam (Adaptive-moment) is a first-order gradient descent optimization based on adaptive estimates of lower-order moments~\cite{Adam}, while AMSGrad is a variant of Adam that enforces the learning rate matrix to be decreasing~\cite{AMSGrad}. The construction of the six DNNs, the choice of optimizers and the execution of the training loop are done using the Chainer framework~\cite{Chainer}. All the codes and the specific parameters used in this study are publicly available in the online repository in~\cite{MyGithub}.

The performance of the six DNN models considered in this study are shown in Figure~\ref{fig:training}. 
The DNN model, with the optimizer and architecture used, are indicated on the top label of each plot.
The accuracy of inference is defined as the number of correct inference divided by the number of input items.
The fluctuating inference accuracy on the testing set (orange lines) is due to the stochasticity of the mini-batch system.
In some cases, the testing accuracy will deviate from the training set performance after some finite training epochs. This behavior is a sign of overfitting wherein the DNN is just memorizing the training dataset. Overfitting can be prevented by reducing the size of DNN. Now, the fact that the testing performance remains close to the training is an indication that the DNN is not just memorizing the training dataset. This result implies that our DNN models achieved generalization within the training S-matrix parametrization.

\hypertarget{sep}{%
	\section{Validation and application}\label{sep}}
\hspace{\parindent}
It is fair to inquire whether the trained DNN is still useful even if the input cross-sections are not generated from the same training S-matrix parametrization in Eq.~\eqref{eq:smat}. We can address this inquiry by producing an independent set of cross-sections using a different S-matrix model and then validating the inference of the trained DNNs. This validation step will allow us to assess the generalization of the DNN beyond the training dataset and will also increase our confidence when the DNN is applied to analyze the experimental data. In this section we use two types of cross-sections generated from the separable potential model. First, the s-wave projected separable potential with constant coupling strength given by
$v(p,p')=\zeta\Lambda^4(p^2+\Lambda^2)^{-1}(p'^2+\Lambda^2)^{-1}$
where $\Lambda$ is the cut-off parameter associated with the Yamaguchi form-factor while $\zeta$ is the coupling strength~\cite{PearceGibson}. By solving the Lippmann-Schwinger equation, we get the elastic S-matrix  
\begin{equation}
	S_{\text{ind}}(p)=\left(\dfrac{p+i\Lambda}{p-i\Lambda}\right)^2
	\left[\dfrac{2(p-i\Lambda)-\zeta\pi\mu\Lambda^3}
	{2(p+i\Lambda)-\zeta\pi\mu\Lambda^3}\right]
	\label{eq:sep_ind}
\end{equation}
with $S(p)=1+ipf(p)$. By inspection, one can produce exactly one isolated pole near the threshold. We call Eq.~\eqref{eq:sep_ind} as the energy-independent coupling S-matrix model.

The second validation cross-sections are to be generated by replacing the parameter $\zeta\rightarrow (E-M_{\text{sep}})\zeta$. The new parameter $M_{sep}$ represents the zero of amplitude. This replacement will give an extra structure to the cross-sections. The S-matrix takes the form
\begin{equation}
	S_{\text{dep}}(p)=\left(\dfrac{p+i\Lambda}{p-i\Lambda}\right)^2
	\left[\dfrac{2(p-i\Lambda)-\zeta\pi\mu\Lambda^3(E-M_{\text{sep}})}
	{2(p+i\Lambda)-\zeta\pi\mu\Lambda^3(E-M_{\text{sep}})}\right].
	\label{eq:sep_dep}
\end{equation}
We call Eq.~\ref{eq:sep_dep} as the energy-dependent coupling S-matrix model. Note that, bound state pole can only be generated by choosing $M_{\text{sep}}<0$; otherwise, we only get resonances.  Inspection of the denominator of S-matrix tells us that a near-threshold pole (bound or virtual) is always accompanied by a nearby simple background pole. This extra structure will help the DNN to distinguish the bound-virtual enhancements

The essential point here is the difference between the background of training S-matrix in Eq.~\eqref{eq:smat} and the separable potential models in Eq.~\eqref{eq:sep_ind} and Eq.~\eqref{eq:sep_dep}. The former has a branch-cut singularity, while the latter is a second-order pole. We suspect that the accuracy of inference will largely depend on the background singularity.  To explore this hypothesis, we produced several cross-sections using the separable potential model with different cut-off parameters and feed them directly to the trained DNN models. The performance is measured by counting the number of correct inferences as we vary the range of cut-off parameters $\Lambda$. 
\begin{figure}
	\centering
	\includegraphics[width=\columnwidth]{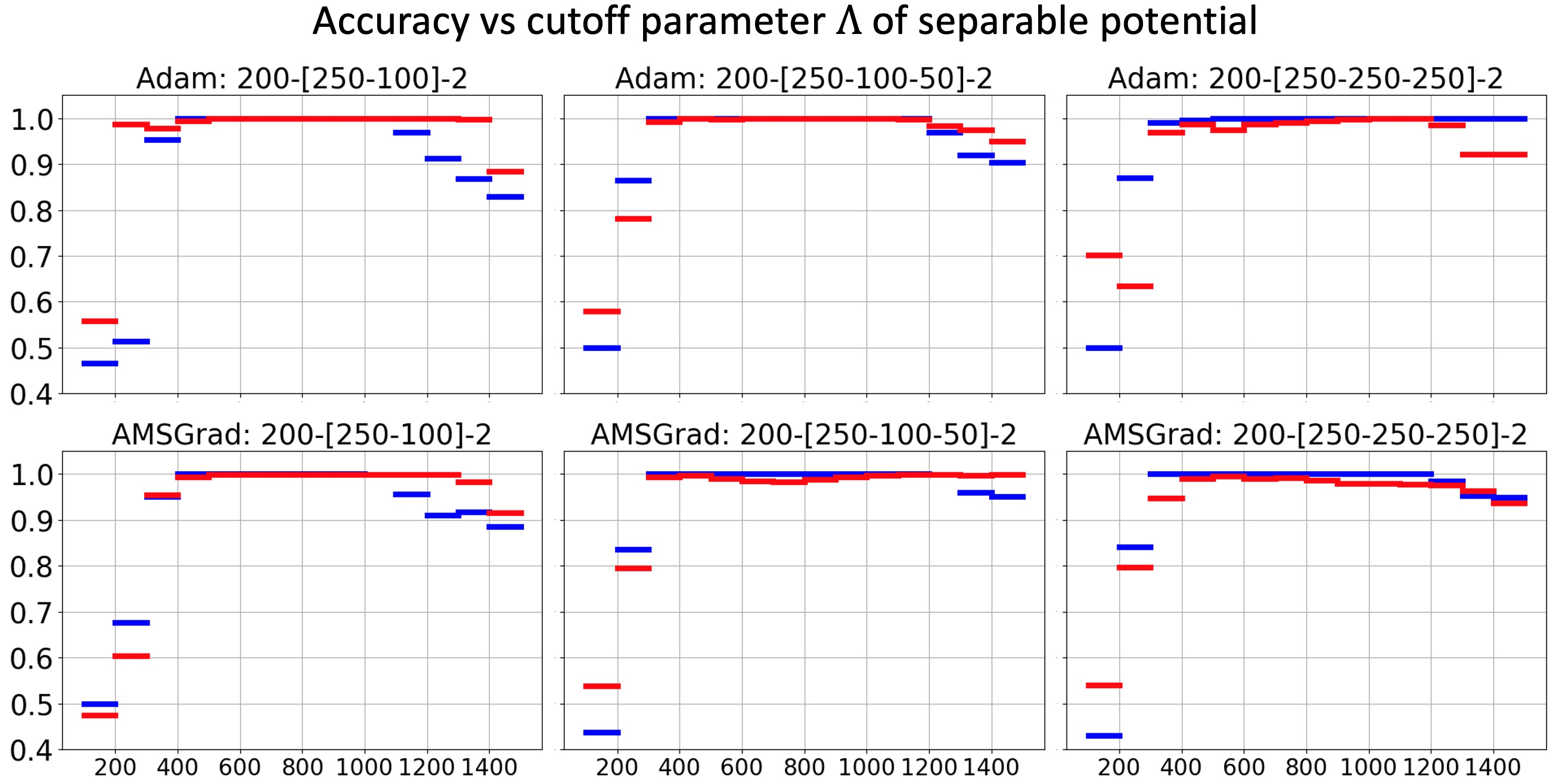}
	\caption{DNN inferences on single-channel cross-sections of separable potential models. Each horizontal line contains $100,000$ cross-sections with blue lines corresponding to energy-independent coupling and red lines to energy-dependent.}
	\label{fig:validate}
\end{figure}

We show the results of our numerical experimentation in Fig.~\ref{fig:validate}. Each horizontal line contains $1\times 10^5$ cross-sections with cut-off range specified by the length of the line segment. Notice that we get poor inference accuracies in the low cut-off regime for all the DNN models considered. We believe that the poor performance is due to the difference between the background singularities in the training and the validation separable S-matrix. This difference is highly emphasized in the small cut-off region because the background singularity here is too close to the threshold. This means that the actual analytic structure of S-matrix is important if the background singularities are too close to the threshold. The low inference accuracy observed in this numerical experiment is unlikely to happen in the actual experimental data because reasonable cut-off parameter starts around $500$ MeV.

High inference accuracies are observed when the cut-off is within the range of $400-1,200\text{ MeV}$. Here, the difference in the background singularities makes no difference on the inference result. This observation justifies that it is sufficient to simulate the background singularities using some convenient parametrization because their exact nature is not relevant in the classification task. As we go above the $1,200\text{ MeV}$ cut-off, the inference accuracy starts to drop. A high cut-off means that the background is too far from the threshold, making it less relevant in the shape of the cross-section in the scattering region. Distinguishing between virtual and bound state pole enhancement with weak background is difficult because the cross-sections of bound and virtual enhancements are almost identical.

\begin{figure}
	\centering
	\includegraphics[width=\columnwidth]{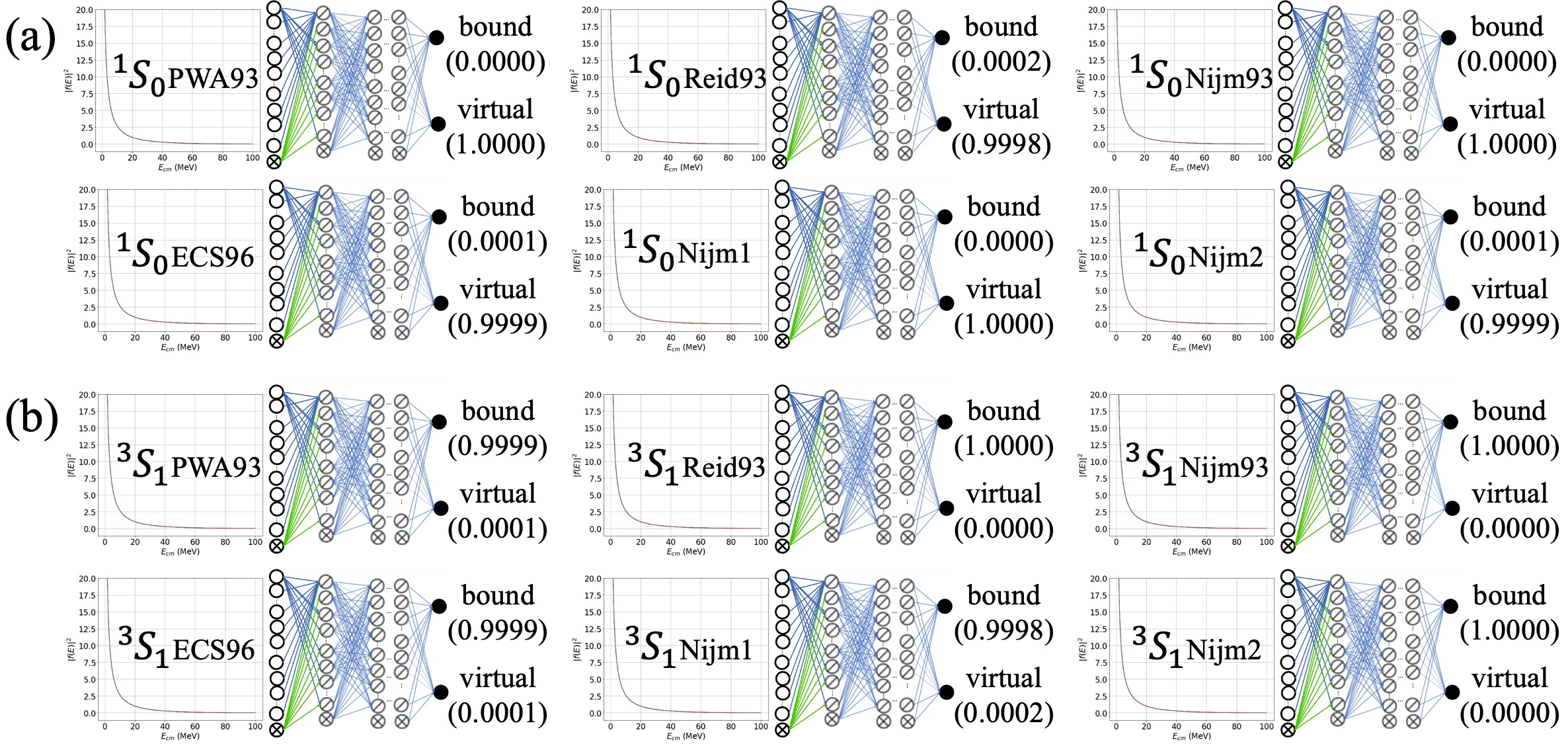}
	\caption{Sample inference output of one of the trained DNN model (AMSGrad:200-[250-250-250]-2) given the (a) ${}^1S_0$ and the (b) ${}^3S_1$ nucleon-nucleon cross-sections as inputs.}
	\label{fig:inference_nn}
\end{figure}

Our experimentation with the separable potential cross-sections guarantee that our trained DNNs are capable of generalization outside the training dataset. The exact nature of background singularity is only relevant if it is very close to the threshold. For reasonable range of cut-off parameters, the DNN inference can be reliable. For the last step, we are now ready to deploy our models to infer the nature of enhancements in the nucleon-nucleon cross-sections. We treat the six partial-wave fits and potential models as our experimental data and feed the corresponding cross-sections directly to our trained DNN. All of our models inferred the correct nature of the  nucleon-nucleon threshold enhancements. That is, ${}^3S_1$ enhancements are identified as due to bound state pole while the ${}^1S_0$ as due to virtual state pole. Figure~\ref{fig:inference_nn} shows one of the sample inference outputs of one of the DNN models for all the nucleon-nucleon cross-sections considered in this study. We see that largest weight is given to the output node representing the correct nature of threshold enhancement. 

The result of our demonstration is simple but its implication can be profound. In this study, we have exploited the unbalanced distribution of singularities of the S-matrix by using two simple backgrounds. The subtle effect of the background in the scattering region is utilized by the DNN as we optimize the weights and biases in the training. We have shown that the background singularity is essential but its exact nature is not relevant in distinguishing the bound-virtual enhancements. Thus, one is justified to simulate the background using some convenient parametrization without compromising the important physics. 

\hypertarget{conc}{%
	\section{Summary}\label{conc}}
\hspace{\parindent}
We have demonstrated that it is possible to distinguish threshold enhancement in the scattering cross-sections using deep learning. The ability of the DNN model to generalize beyond the training dataset is shown using the cross-sections generated from the separable potential in the validation stage. It turns out that the background singularities such as branch cut or distant poles are sufficient for the DNN model to distinguish virtual and bound state enhancements at the threshold. Finally, we have shown that using only the scattering cross-sections, it is possible to identify the existence of nucleon-nucleon bound state.

\section*{Acknowledgment}
This study is supported in part by JSPS KAKENHI Grant No. JP17K14287, and by MEXT as “Priority Issue on Post-K computer” (Elucidation of the Fundamental Laws and Evolution of the Universe) and SPIRE (Strategic Program for Innovative Research).
DLBS is supported in part by the UP OVPAA FRASDP and DOST-PCIEERD postdoctoral research grant. YI is partly supported by JSPS KAKENHI Nos. JP17K14287 (B) and 21K03555 (C).
AH is supported in part by JSPS KAKENHI No. JP17K05441 (C) and Grants-in-Aid for Scientific Research on Innovative Areas, No. 18H05407 and 19H05104.

%\bibliography{mybib}

\end{document}